\documentclass[symmetry,article,accept,moreauthors,dvipdfm,12pt,a4paper]{mdpi} 
\lastpage{1484}
\doinum{10.3390/sym2031461}
\pubvolume{2}
\pubyear{2010}
\history{Received: 9 June 2010; in revised form: 8 July 2010 / Accepted: 9 July 2010 / \linebreak Published: 12 July 2010}

\usepackage{amssymb}

\usepackage{hyphenat}
\usepackage[sort&compress]{natbib}
\usepackage{hypernat}

\def\ii{{\rm i}}
\def\FF{{\bf F}}
\def\dd{{\rm d}}
\def\onehalf{\textstyle\frac12}
\def\onefour{\textstyle\frac14}
\def\be{\begin{equation}}
\def\ee{\end{equation}}

\Title{SU(2) and SU(1,1) Approaches to Phase Operators and
Temporally Stable Phase States: Applications to Mutually Unbiased
Bases and Discrete Fourier Transforms}

\Author{Natig M. Atakishiyev $^{1}$, Maurice R. Kibler $^{2,3,4,\star}$ and Kurt Bernardo Wolf $^{5}$}

\address{%
$^{1}$ Instituto de Matem\'aticas, Universidad Nacional Aut\'onoma de M\'exico, Av.\ Universidad s/n,
Cuernavaca, Morelos 62251, Mexico\\
$^{2}$ Universit\'e de Lyon, 37 rue du repos, 69361 Lyon, France\\
$^{3}$ Universit\'e Claude Bernard and CNRS/IN2P3, 43 Bd du 11 Novembre 1918, F-69622
\linebreak Villeurbanne, France\\
$^{4}$ Institut de Physique Nucl\'eaire, 43 Bd du 11 Novembre 1918, F-69622 Villeurbanne, France\\
$^{5}$ Instituto de Ciencias F\'{\i}sicas, Universidad Nacional Aut\'onoma de M\'exico, Av.\ Universidad s/n,
Cuernavaca, Morelos 62251, Mexico
}

\corres{E-Mail: m.kibler@ipnl.in2p3.fr; \linebreak Tel.: 33 (0)4 72 44 82 35.}

\abstract{We propose a group-theoretical approach to the generalized
oscillator algebra ${\cal A}_{\kappa}$ recently investigated in {\em
J.\ Phys.\ A: Math.\ Theor.} {\bf 2010}, {\em 43}, 115303. The case
$\kappa \ge 0$ corresponds to the noncompact group SU(1,1) (as for
the harmonic oscillator and the P\" oschl-Teller systems) while the
case $\kappa < 0$ is described by the compact group SU(2) (as for
the Morse system). We construct the phase operators and the
corresponding  temporally stable phase eigenstates for ${\cal
A}_{\kappa}$ in this group-theoretical context. The SU(2) case is
exploited for deriving families of mutually unbiased bases used in
quantum information. Along this vein, we examine some
characteristics of a quadratic discrete Fourier transform in
connection with generalized quadratic Gauss sums and generalized
Hadamard matrices. }



\begin{document}

\section{Introduction}
The use of a generalized oscillator algebra for characterizing a
dynamical system gave rise to a great deal of papers. Among many
works, we may quote the polynomial Heisenberg algebra worked out in
the context of supersymmetry
\cite{Fernandez-1,Fernandez-2,Fernandez-3}, the deformed Heisenberg
algebra introduced in connection with parafermionic and parabosonic
systems \cite{Plyu96,Plyu97,Plyu00,Plyu10}, the
$C_{\lambda}$-extended oscillator algebra developed in the framework
of parasupersymmetric quantum mechanics
\cite{Ques1,Ques2,Ques3,Ques4}, and the generalized Weyl-Heisenberg
algebra $W_k$ related to $\mathbb{Z}_k$--graded supersymmetric
quantum mechanics \cite{Kib1,Kib2,Kib3,Kib4,DaoKib2010}. In this
direction, the construction of a truncated generalized oscillator
algebra was developed by several authors. In particular, the pioneer
work along this line by Pegg and Barnett led to calculating the
phase properties of the electromagnetic field \cite{Pegg}. Let us
also mention the works \cite{Roy1,Roy2} in relation with orthogonal
polynomials of a discrete variable and \cite{DaoKib2010} in
connection with phase operators and dynamical systems.

Recently, a generalized oscillator algebra ${\cal A}_{\kappa}$, a
one-parameter algebra that is a particular case of the algebra
$W_1$, was studied for the purpose of defining phase operators and
the corresponding phase eigenstates \cite{DaoKib2010}. In addition,
it was shown that the phase states for ${\cal A}_{\kappa}$ with
$\kappa > 0$, which are particular coherent states
\cite{Perelomov86,Gazeau}, can serve to construct mutually unbiased
bases which are of considerable interest in quantum information and
quantum computing \cite{DaoKib2010}.

It is the aim of the present paper to analyze the algebra ${\cal
A}_{\kappa}$ from the point of view of group theory. Since ${\cal
A}_{\kappa}$ can describe the Morse system for $\kappa < 0$ as well
as the harmonic oscillator and the P\" oschl-Teller systems for
$\kappa \geq 0$, we expect that the groups SU(2) and SU(1,1) play a
central role. The search for phase operators and temporally stable
phase states thus amounts to study generalized coherent states for
SU(2) and SU(1,1).

The material presented here is organized as follows. Section 2 deals
with the generalized oscillator algebra ${\cal A}_{\kappa}$ and its
connection with the Lie algebra of SU(2) and SU(1,1). The phase
operators and the phase states introduced in \cite{DaoKib2010} are
described in the framework of SU(2) and SU(1,1). Section 4 is
devoted to a truncation of the algebra ${\cal A}_{\kappa}$. In
section 5, the phase operator for the group SU(2) is shown to be of
relevance for the determination of mutually unbiased bases ({\em
cf}.
\cite{WRa,CCCC05,IJMP1,IJMP2,AlbKib1,AlbKib2,Kib-quons,CCCC08,JPhysA08,JPhys09,KibDubna}).
Finally, the quadratic transformation that connects the phase states
for SU(2) to angular momentum states is studied in \linebreak
Section 6. This transformation generalizes the discrete Fourier
transform whose the main properties are given in the appendix.

The notations are standard. Let us simply mention that: $\delta_{a ,
b}$ stands for the Kronecker symbol of $a$ and $b$, $I$ for the
identity operator, $A^{\dagger}$ for the adjoint of the operator
$A$, and $[A , B]$ for the commutator of the operators $A$ and $B$.
The bar indicates complex conjugation and matrices are generally
written with bold-face letters (${\bf I_d}$ is the $d$-dimensional
identity matrix). We use a notation of type $\vert \psi \rangle$ for
a vector in an Hilbert space and we denote $\langle \phi \vert \psi
\rangle$ and $\vert \phi \rangle \langle \psi \vert$ respectively
the inner and outer products of the vectors $\vert \psi \rangle$ and
$\vert \phi \rangle$. As usual $\mathbb{N}$, $\mathbb{N}^*$,
$\mathbb{Z}$ and $\mathbb{R}_+$ are the sets of integers, strictly
positive integers, relative integers and positive real numbers;
$\mathbb{R}$ and $\mathbb{C}$ the real and complex fields; and
$\mathbb{Z}/d\mathbb{Z}$ the ring of \linebreak integers $0, 1,
\ldots, d-1$ modulo $d$.

\section{Generalized Oscillator Algebra}

\subsection{The Algebra}

Following \cite{DaoKib2010}, we start from the algebra ${\cal
A}_{\kappa}$ spanned by the three linear operators $a^-$, $a^+$ and
$N$ satisfying the following commutation relations
      \begin{eqnarray}
[a^- , a^+] = I + 2 \kappa N,       \qquad
[N , a^{\pm}] = \pm a^{\pm},        \qquad
\left( a^- \right)^{\dagger} = a^+, \qquad
N^{\dagger} = N
      \label{thealgebraAkappa}
      \end{eqnarray}
where $\kappa$ is a real parameter. In the particular case $\kappa =
0$, the algebra ${\cal A}_{0}$ is the usual harmonic oscillator
algebra. In the case $\kappa \not= 0$, the operators $a^-$, $a^+$
and $N$ in (\ref{thealgebraAkappa}) generalize the annihilation,
creation and number operators used for the harmonic oscillator.
Thus, the algebra ${\cal A}_{\kappa}$ can be referred to as a
generalized oscillator algebra. In fact, the algebra ${\cal
A}_{\kappa}$ represents a particular case of the generalized
Weyl-Heisenberg algebra $W_k$ introduced in
\cite{Kib1,Kib2,Kib3,Kib4} to describe a fractional supersymmetric
oscillator. A similar algebra, namely the $C_{\lambda}$-extended
oscillator algebra, was studied in connection with a generalized
oscillator \cite{Ques1,Ques2,Ques3,Ques4}.

\subsection{The Oscillator Algebra as a Lie Algebra}

The case $\kappa = 0$ corresponds of course to the usual
Weyl-Heisenberg algebra. It can be shown that the cases $\kappa < 0$
and $\kappa > 0$ considered in \cite{DaoKib2010} are associated with
the Lie algebras of the groups SU(2) and SU(1,1), respectively. We
shall consider in turn the cases when $\kappa < 0$ and $\kappa > 0$.

For $\kappa < 0$, we introduce the operators $J_-$, $J_+$ and $J_3$
defined via
      \begin{eqnarray}
J_- := \frac{1}{\sqrt{-\kappa}} a^-,           \qquad
J_+ := \frac{1}{\sqrt{-\kappa}} a^+,           \qquad
J_3 := \frac{1}{2 \kappa} (I + 2 \kappa N)
      \label{generators SU(2)}
      \end{eqnarray}
They satisfy the commutation relations
      \begin{eqnarray}
[J_+ , J_-] = 2 J_3, \qquad
[J_3 , J_+] =   J_+, \qquad
[J_3 , J_-] = - J_-
      \label{Lie algebra SU(2)}
      \end{eqnarray}
and therefore span the Lie algebra of SU(2).

Similarly for $\kappa > 0$ the operators $K_-$, $K_+$ and $K_3$, given by
      \begin{eqnarray}
K_- := \frac{1}{\sqrt{\kappa}} a^-,           \qquad
K_+ := \frac{1}{\sqrt{\kappa}} a^+,           \qquad
K_3 := \frac{1}{2 \kappa} (I + 2 \kappa N)
      \label{generators SU(1,1)}
      \end{eqnarray}
lead to the Lie brackets
      \begin{eqnarray}
[K_+ , K_-] = - 2 K_3, \qquad
[K_3 , K_+] =   K_+,   \qquad
[K_3 , K_-] = - K_-
      \label{Lie algebra SU(1,1)}
      \end{eqnarray}
of the group SU(1,1).

\subsection{Rotated Shift Operators for Su(2) and Su(1,1)}

We are now in a position to reconsider some of the results of
\cite{DaoKib2010} in terms of the Lie algebras su(2) and su(1,1).
This will shed new  light on the usual treatments of the
representation theory of SU(2) and SU(1,1) as far as the action on
the representation space of the shift operators of these groups
\linebreak are concerned.

Let us first recall that in the generic case ($\kappa \in
\mathbb{R}$), the algebra ${\cal A}_{\kappa}$ admits a Hilbertian
representation for which the operators $a^-$, $a^+$ and $N$ act on a
Hilbert space ${\cal F}_{\kappa}$ spanned by the basis $\{ \vert n
\rangle : n = 0, 1, \ldots \}$ that is orthonormal with respect to
an inner product $\langle n \vert n' \rangle = \delta_{n,n'}$. The
dimension of ${\cal F}_{\kappa}$ is finite when $\kappa < 0$ or
infinite when $\kappa > 0$. The representation is defined through
\cite{DaoKib2010}
      \begin{eqnarray}
& & a^+ \vert n \rangle = \sqrt{F(n+1)} e^{{-\ii [F(n+1)- F(n)]  \varphi }} \vert n+1 \rangle \label{action de aplus sur les n}  \\
& & a^- \vert n \rangle = \sqrt{F(n)}   e^{{+\ii [F(n) - F(n-1)] \varphi }} \vert n-1 \rangle \label{action de amoins sur les n} \\
& & a^- \vert 0 \rangle = 0, \qquad
    N   \vert n \rangle = n \vert n \rangle
    \label{reste action sur les n}
      \end{eqnarray}
where $\varphi$ is an arbitrary real parameter and the function
$F : \mathbb{N} \to \mathbb{R}_+$ satisfies
      \begin{eqnarray}
F(n+1) - F(n) = 1 + 2 \kappa n, \qquad F(0) = 0  \ \ \Rightarrow  \ \ F(n) = n [1 + \kappa (n - 1)]
    \label{recurrence}
      \end{eqnarray}
Obviously, for $\kappa > 0$ the dimension of ${\cal F}_{\kappa}$ is
infinite. In contrast, for $\kappa < 0$ the space ${\cal
F}_{\kappa}$ is finite-dimensional with a dimension given by
      \begin{eqnarray}
d := 1 - \frac{1}{\kappa}  \ \ {\rm with}  \ \ - \frac{1}{\kappa} \in \mathbb{N}^*
    \label{dimension}
      \end{eqnarray}
It is thus possible to transcribe (\ref{action de aplus sur les
n})-(\ref{reste action sur les n}) in terms of the Lie algebras
su(2) and su(1,1).

\subsubsection{2.3.1. The Su(2) Case}
Let us consider the ($2j+1$)-dimensional irreducible representation
of SU(2) spanned by the orthonormal set
      \begin{eqnarray}
B_{2j+1} := \{ \vert j , m \rangle : m = j, j-1, \ldots, -j \}
    \label{canonical basis}
      \end{eqnarray}
where $\vert j , m \rangle$ is an eigenvector
of $J_3$ and of the Casimir operator
      \begin{eqnarray}
J^2 := J_+ J_- + J_3(J_3 - 1)
    \label{Casimir SU2}
      \end{eqnarray}
We know that
      \begin{eqnarray}
J^2 \vert j , m \rangle = j(j+1) \vert j , m \rangle, \qquad J_3 \vert j , m \rangle = m \vert j , m \rangle
    \label{eigenequations SU2}
      \end{eqnarray}
with $m = j, j-1, \ldots, -j$ for fixed $j$ ($2j \in \mathbb{N}$).
Following
\cite{IJMP2,AlbKib1,AlbKib2,Kib-quons,CCCC08,JPhysA08,JPhys09,KibDubna},
we make the identifications
      \begin{eqnarray}
\vert n \rangle \leftrightarrow \vert j , m \rangle, \qquad n \leftrightarrow j+m
    \label{identification SU2}
      \end{eqnarray}
Consequently, we have
      \begin{eqnarray}
d = 2 j + 1 = 1 - \frac{1}{\kappa}
    \label{identification dimensions SU2}
      \end{eqnarray}
which leads to the relation
      \begin{eqnarray}
2 j \kappa = -1  \ \ \Leftrightarrow  \ \ - \frac{1}{\kappa} = 2j
    \label{relation jkappa}
      \end{eqnarray}
that is crucial for the connection between ${\cal A}_{\kappa}$ and
su(2). It is to be noted that (\ref{identification
SU2})--(\ref{relation jkappa}) are compatible with (\ref{generators
SU(2)}). We can then rewrite (\ref{action de aplus sur les n}) and
(\ref{action de amoins sur les n}) in the su(2) framework. In fact,
by combining (\ref{generators SU(2)}), (\ref{recurrence}) and
(\ref{relation jkappa}) with (\ref{action de aplus sur les n}) and
(\ref{action de amoins sur les n}), we obtain
      \begin{eqnarray}
& & J_+ \vert j,m \rangle = \sqrt{(j-m)(j+m+1)} e^{ - 2 \ii m     \kappa \varphi } \vert j,m+1 \rangle \label{action de Jplus}  \\
& & J_- \vert j,m \rangle = \sqrt{(j+m)(j-m+1)} e^{   2 \ii (m-1) \kappa \varphi } \vert j,m-1 \rangle \label{action de Jmoins}
      \end{eqnarray}
Equations (\ref{action de Jplus}) and (\ref{action de Jmoins})
differ from the usual relations, well known in angular momentum
theory, by the introduction of the phase factor $\varphi$. The
standard relations, that correspond to the Condon-Shortley phase
convention of atomic spectroscopy, are recovered when $\varphi = 0$.

Although there is no interdiction to have $\varphi \not= 0$, it is
worthwhile to look for the significance of the introduction of
$\varphi$. Let us call $\hat{J}_+$ and $\hat{J}_-$ those operators
${J}_+$ and ${J}_-$ which correspond to $\varphi = 0$, respectively.
It is easy to show then that $\hat{J}_{\pm}$ and ${J}_{\pm}$ are
connected by the similarity transformation
      \begin{eqnarray}
\hat{J}_{\pm} = e^{-\ii X \kappa \varphi} {J}_{\pm} e^{\ii X \kappa \varphi}
    \label{rotated J for SU2}
      \end{eqnarray}
where the operator $X$ reads
      \begin{eqnarray}
X := J^2 - J_3(J_3 - 1) = {J}_+ {J}_- = \hat{J}_+ \hat{J}_-
    \label{definition de X}
      \end{eqnarray}
Note that the nonlinear transformation ${J}_{\pm} \leftrightarrow
\hat{J}_{\pm}$, defined by (\ref{rotated J for SU2}), leaves
invariant the Casimir operator $J^2$ of SU(2). We shall see in
section 5 that the parameter $\varphi$ is essential in order to
generate mutually unbiased bases.

\subsubsection{2.3.2. The Su(1,1) Case}
The representation theory of SU(1,1) is well known (see for example
\cite{Perelomov86}). We shall be concerned here with the positive
discrete series $D_+'$ of SU(1,1). The representation associated
with the Bargmann index $k$ can be defined via
      \begin{eqnarray}
& & K_+ \vert k, k+n \rangle = \sqrt{(2k+n)  (n+1)} e^{ - \ii \psi(k,n)   } \vert k,k+n+1 \rangle \label{actionclas de Kplus}  \\
& & K_- \vert k, k+n \rangle = \sqrt{(2k+n-1) n   } e^{   \ii \psi(k,n-1) } \vert k,k+n-1 \rangle \label{actionclas de Kmoins} \\
& & K_3 \vert k, k+n \rangle = (k+n) \vert k, k+n \rangle
    \label{actionclas de K3}
      \end{eqnarray}
with
      \begin{eqnarray}
K^2 \vert k, k+n \rangle = k(1 - k) \vert k,k+n \rangle, \qquad K^2 := K_+ K_- - K_3 (K_3 - 1)
    \label{Casimir SU11}
      \end{eqnarray}
where $n \in \mathbb{N}$ and $K^2$ stands for the Casimir operator
of SU(1,1). This infinite-dimensional representation is spanned by
the orthonormal set $\{ \vert k, k+n \rangle : n \in \mathbb{N} \}$.
Equations (\ref{actionclas de Kplus}) and (\ref{actionclas de
Kmoins}) differ from the standard relations \cite{Perelomov86} by
the introduction of the real-valued phase function $\psi$. Such a
function is introduced, in a way paralleling the introduction of the
phase factors in (\ref{action de aplus sur les n}) and (\ref{action
de amoins sur les n}), to make precise the connection between ${\cal
A}_{\kappa}$ and su(1,1) for $\kappa > 0$. The relative phases in
(\ref{actionclas de Kplus}) and (\ref{actionclas de Kmoins}) are
such that $K_+$ is the adjoint of $K_-$. For fixed $\kappa$ and $k$,
we make the identification
      \begin{eqnarray}
\vert n \rangle \leftrightarrow \vert k, k+n \rangle
    \label{identification SU11}
      \end{eqnarray}
Then, from (\ref{generators SU(1,1)}) we get the central relation
      \begin{eqnarray}
2 k \kappa = 1  \ \ \Leftrightarrow  \ \ \frac{1}{\kappa} = 2k
    \label{relation kkappa}
      \end{eqnarray}
to be compared with (\ref{relation jkappa}). Furthermore, by
combining (\ref{generators SU(1,1)}), (\ref{action de aplus sur les
n}), (\ref{action de amoins sur les n}), (\ref{actionclas de
Kplus}), (\ref{actionclas de Kmoins}), (\ref{identification SU11})
and (\ref{relation kkappa}) we get
      \begin{eqnarray}
F(n) = n \left[ 1 + \frac{1}{2k}(n-1) \right], \qquad \psi(k,n  ) = \frac{1}{k} (k + n) \varphi
    \label{relation F et psi}
      \end{eqnarray}
Finally, the action of the shift operators $K_+$ and $K_-$ on a generic vector $\vert k, k+n \rangle$
can be rewritten as
      \begin{eqnarray}
& & K_+ \vert k, k+n \rangle = \sqrt{(2k+n)  (n+1)} e^{ -2 \ii (k+n)   \kappa \varphi } \vert k,k+n+1 \rangle \label{actionclas de Kplusbis}  \\
& & K_- \vert k, k+n \rangle = \sqrt{(2k+n-1) n   } e^{  2 \ii (k+n-1) \kappa \varphi } \vert k,k+n-1 \rangle \label{actionclas de Kmoinsbis}
      \end{eqnarray}
The particular case $\varphi = 0$ in (\ref{actionclas de Kplusbis})
and (\ref{actionclas de Kmoinsbis}) gives back the standard
relations for SU(1,1).

The operators ${K}_+$ and ${K}_-$ are connected to the operators
$\hat{K}_+$ and $\hat{K}_-$ corresponding to $\varphi = 0$ by
      \begin{eqnarray}
\hat{K}_{\pm} = e^{\ii Y \kappa \varphi} {K}_{\pm} e^{-\ii Y \kappa \varphi}
    \label{rotated K for SU11}
      \end{eqnarray}
with
      \begin{eqnarray}
Y := K^2 + K_3(K_3 - 1) = {K}_+ {K}_- = \hat{K}_+ \hat{K}_-
    \label{definition de Y}
      \end{eqnarray}
so that the nonlinear transformation ${K}_{\pm} \leftrightarrow
\hat{K}_{\pm}$, defined by (\ref{rotated K for SU11}), leaves
invariant the Casimir operator $K^2$ of SU(1,1).

\section{Phase Operators}
Phase operators were defined in \cite{DaoKib2010} from a
factorization of the annihilation operator $a^-$ of ${\cal
A}_{\kappa}$. We shall transcribe this factorization in terms of the
lowering generators $J_-$ and $K_-$ of SU(2) and SU(1,1),
respectively.

\subsection{The Su(2) Case}
Let us define $E_d$ via
      \begin{eqnarray}
J_- = E_d \sqrt{J_+ J_-}
    \label{Jm en polaire}
      \end{eqnarray}
The operator $E_d$ can be developed as
      \begin{eqnarray}
E_d = \sum_{m=-j}^{j} e^{2 \ii (m-1) \kappa \varphi} \vert j , m-1 \rangle \langle j,m \vert
    \label{def de Ed}
      \end{eqnarray}
where $m-1$ should be understood as $j$ when $m = -j$. Consequently
      \begin{eqnarray}
E_d \vert j , m \rangle = e^{2 \ii (m-1) \kappa \varphi} \vert j , m-1 \rangle  \ \ {\rm for}  \ \ m \not= -j
    \label{action1 de Ed}
      \end{eqnarray}
and
      \begin{eqnarray}
E_d \vert j , -j \rangle = e^{-\ii \varphi} \vert j , j \rangle  \ \ {\rm for}  \ \ m = -j
    \label{action2 de Ed}
      \end{eqnarray}
It is clear from (\ref{action1 de Ed}) and (\ref{action2 de Ed})
that the operator $E_d$ is unitary.

In order to show that $E_d$ is a phase operator, we consider the eigenvalue equation
      \begin{eqnarray}
E_d \vert z \rangle = z \vert z \rangle, \qquad
\vert z \rangle := \sum_{m = -j}^{j} d_m z^{j+m} \vert j,m \rangle, \qquad
z \in {\mathbb{C}}, \qquad
d_m \in {\mathbb{C}}
      \label{eigenequation for Ed}
      \end{eqnarray}
It can be shown that the determination of normalized eigenstates
$\vert z \rangle$ satisfying (\ref{eigenequation for Ed}) requires
that the condition
      \begin{eqnarray}
z^{2j+1} = 1
      \label{condition sur z}
      \end{eqnarray}
be fulfilled. Hence, the complex variable $z$ is a root of unity given by
     \begin{eqnarray}
z = q^{\alpha}, \qquad
q = e^{2 \pi \ii / (2j+1)}, \qquad
\alpha = 0, 1, \ldots, 2j
      \label{valeurs de z}
      \end{eqnarray}
As a result, the states $\vert z \rangle$ depend on a continuous
parameter $\varphi$ and a discrete parameter $\alpha$. They shall be
written as $\vert \varphi , \alpha \rangle$. A lengthy calculation
leads to
      \begin{eqnarray}
\vert z \rangle \equiv \vert \varphi , \alpha \rangle = \frac{1}{\sqrt{2j+1}} \sum_{m = -j}^{j}
e^{\ii (j+m) (j-m+1) \kappa \varphi} q^{\alpha (j+m)} \vert j,m \rangle
      \label{state varphi alpha}
      \end{eqnarray}
The latter states satisfy
     \begin{eqnarray}
E_d \vert \varphi , \alpha \rangle = q^{\alpha} \vert \varphi , \alpha \rangle = e^{2 \pi \ii \alpha/(2j+1)} \vert \varphi , \alpha \rangle,
\qquad \alpha = 0, 1, \ldots, 2j
      \label{eigenvalue Ed}
      \end{eqnarray}
Thus, the states $\vert \varphi , \alpha \rangle$ are phase states
and the unitary operator $E_d$ is a phase operator, with a
non-degenerate spectrum, associated with SU(2). Furthermore, the
eigenvectors of $E_d$ satisfy
      \begin{eqnarray}
      U(t) \vert \varphi , \alpha \rangle = \vert \varphi + t, \alpha \rangle
      \label{tempor stable en J}
      \end{eqnarray}
where
      \begin{eqnarray}
      U(t) := e^{-\ii H t}, \qquad H := - \kappa X = - \kappa J_+ J_-
      \label{operator U en J}
      \end{eqnarray}
and $t$ is a real parameter. Equation (\ref{tempor stable en J})
indicates that the phase states $\vert \varphi , \alpha \rangle$ are
temporally stable, an important property to determine the so-called
mutually unbiased bases \cite{DaoKib2010}. Note that they are not
all orthogonal (the states with the same $\varphi$ are of course
orthogonal) and they satisfy the closure property
      \begin{eqnarray}
\sum_{\alpha = 0}^{2j} \vert \varphi , \alpha \rangle \langle \varphi , \alpha \vert = I
      \end{eqnarray}
for fixed $\varphi$ (see also \cite{DaoKib2010}).

\subsection{The Su(1,1) Case}
By writing
      \begin{eqnarray}
K_- = E_{\infty} \sqrt{K_+ K_-}
    \label{Km en polaire}
      \end{eqnarray}
it can be shown that
      \begin{eqnarray}
E_{\infty} = \sum_{n=0}^{\infty} e^{2 \ii (k+n) \kappa \varphi} \vert k , k+n \rangle \langle k , k+n+1 \vert
    \label{def de Einfini}
      \end{eqnarray}
The operator $E_{\infty}$ has the following property
      \begin{eqnarray}
E_{\infty} (E_{\infty})^{\dagger} = (E_{\infty})^{\dagger} E_{\infty} + \vert k,k \rangle \langle k,k \vert = I
    \label{pseudo unitarity of Einfini}
      \end{eqnarray}
Thus, it is not unitary in contrast with the case of the operator
$E_d$ for su(2).

Let us look for normalized states $\vert z \rangle$ such that
      \begin{eqnarray}
E_{\infty} \vert z \rangle = z \vert z \rangle, \qquad
\vert z \rangle := \sum_{n=0}^{\infty} c_n z^{n} \vert k,k+n \rangle, \qquad
z \in {\mathbb{C}}, \qquad
c_n \in {\mathbb{C}}
      \label{eigenequation for Einfinifi}
      \end{eqnarray}
One readily finds that
       \begin{eqnarray}
\vert z \rangle = \sqrt{1 - |z|^2} \sum_{n=0}^{\infty} z^n e^{- \ii n(2k+n-1) \kappa \varphi} \vert k,k+n \rangle,
\qquad |z| < 1
      \label{with square root}
      \end{eqnarray}
up to a phase factor. Following \cite{VouBM96} and
\cite{DaoKib2010}, we define the states $\vert \varphi , \theta
\rangle$ by
       \begin{eqnarray}
 \vert \varphi , \theta \rangle := \lim_{z \rightarrow e^{\ii\theta}} \frac{1}{\sqrt{1 - |z|^2}} \vert z \rangle
      \label{state theta varphi}
       \end{eqnarray}
where $\theta \in [-\pi , +\pi[$. One thus obtains that
      \begin{eqnarray}
\vert \varphi , \theta \rangle = \sum_{n=0}^{\infty} e^{\ii n \theta} e^{- \ii n (2k+n-1) \kappa \varphi} \vert k,k+n \rangle
      \label{state theta varphi as a series}
      \end{eqnarray}
The states (\ref{state theta varphi as a series}), defined on the
unit circle $S^1$, have the property
      \begin{eqnarray}
E_{\infty} \vert \varphi , \theta \rangle  = e^{\ii \theta} \vert \varphi , \theta \rangle,
\qquad - \pi \leq \theta < \pi
      \label{propert of Einfini}
      \end{eqnarray}
The operator $E_{\infty}$ is thus a nonunitary phase operator
associated with SU(1,1). As a particular case of the phase states
$\vert \varphi , \theta \rangle$, the states $\vert 0 , \theta
\rangle$ corresponding to $\varphi = 0$ are identical to the phase
states introduced in \cite{VouBM96} for SU(1,1). The parameter
$\varphi$ ensures that the states $\vert \varphi , \theta \rangle$
are temporally stable with respect to
      \begin{eqnarray}
      U(t) := e^{-\ii H t}, \qquad H := \kappa Y = \kappa K_+ K_-
      \label{operator U en K}
      \end{eqnarray}
in the sense that
      \begin{eqnarray}
      U(t) \vert \varphi , \theta \rangle = \vert \varphi + t , \theta \rangle
      \label{tempor stable en K}
      \end{eqnarray}
for any real value of $t$. Note that, for fixed $\varphi$, the phase
states $\vert \varphi , \theta \rangle$, satisfy the closure relation
      \begin{eqnarray}
\frac{1}{2\pi} \int_{-\pi}^{+\pi} d\theta \vert \varphi , \theta \rangle \langle \varphi , \theta \vert = I
      \end{eqnarray}
but they are neither normalized nor orthogonal.

\section{Truncated Generalized Oscillator Algebra}

The idea of a truncated algebra for the harmonic oscillator goes
back to Pegg and Barnett \cite{Pegg}. Truncated algebras for
generalized oscillators were introduced in
\cite{DaoKib2010,Roy1,Roy2}. In \cite{DaoKib2010}, a truncated
oscillator algebra ${\cal A}_{\kappa,s}$ associated with the algebra
${\cal A}_{\kappa}$ was considered both in the infinite-dimensional
case ($\kappa \geq 0$) and the finite-dimensional case ($\kappa <
0$). The introduction of such a truncated algebra makes it possible
to define a unitary phase operator for $\kappa \geq 0$ and to avoid
degeneracy problems for $\kappa < 0$. We shall briefly revisit in
this section the truncation of the generalized oscillator algebra
${\cal A}_{\kappa}$ in an approach that renders more precise the
relationship between ${\cal A}_{\kappa, s}$ and ${\cal A}_{\kappa}$.

Let us start with the two operators
   \begin{eqnarray}
c^+ &=& a^+ - \sum_{n=s}^{d(\kappa)} \sqrt{F(n)} e^{ - \ii [F(n) - F(n-1)] \varphi } \vert n \rangle \langle n-1 \vert
   \label{cplus} \\
c^- &=& a^- - \sum_{n=s}^{d(\kappa)} \sqrt{F(n)} e^{ + \ii [F(n) - F(n-1)] \varphi } \vert n-1 \rangle \langle n \vert
   \label{cmoins}
   \end{eqnarray}
where $d(\kappa) = d-1$ or $\infty$ according to whether $\kappa <
0$ or $\kappa \geq 0$. The finite truncation index $s$ is arbitrary
for $\kappa \geq 0$ and less than $d$ for $\kappa < 0$. It is
straightforward to prove that
      \begin{eqnarray}
& & c^+ \vert n \rangle = \sqrt{F(n+1)} e^{{-\ii [F(n+1)- F(n)]  \varphi }} \vert n+1 \rangle  \ \ {\rm for}  \ \ n = 0, 1, \ldots, s-2
\label{action1 de cplus sur les n}  \\
& & c^+ \vert n \rangle = 0  \ \ {\rm for}  \ \ n = s-1, s, \ldots, d(\kappa)
\label{action2 de cplus sur les n}  \\
& & c^- \vert n \rangle = \sqrt{F(n)}   e^{{+\ii [F(n) - F(n-1)] \varphi }} \vert n-1 \rangle  \ \ {\rm for}  \ \ n = 1, 2, \ldots, s-1
\label{action1 de cmoins sur les n} \\
& & c^- \vert n \rangle = 0  \ \ {\rm for}  \ \ n=0  \ \ {\rm and}  \ \ n = s, s+1, \ldots, d(\kappa)
\label{action2 de cmoins sur les n}
      \end{eqnarray}
Therefore, the operators $c^-$ and $c^+ = (c^-)^{\dagger}$ lead to
the null vector when acting on the vectors of the space ${\cal
F}_{\kappa}$ that do not belong to its subspace ${\cal
F}_{\kappa,s}$ spanned by the set $\{ \vert 0 \rangle, \vert 1
\rangle, \ldots, \vert s-1 \rangle \}$. In this sense, $c^+$ and
$c^-$ differ from the operators $b^+$ and $b^-$ of
\cite{DaoKib2010}.

In the light of Equations (\ref{action1 de cplus sur les
n})--(\ref{action2 de cmoins sur les n}), the passage from the
algebra ${\cal A}_{\kappa}$ to the truncated algebra ${\cal
A}_{\kappa, s}$ should be understood as the restriction of the space
${\cal F}_{\kappa}$ to its subspace ${\cal F}_{\kappa,s}$ together
with the replacement of the commutation relations in
(\ref{thealgebraAkappa}) by
      \begin{eqnarray}
[c^- , c^+] = I + 2 \kappa N - F(s) \vert s-1 \rangle \langle s-1 \vert - \sum_{n=s}^{d(\kappa)} (1 + 2 \kappa n) \vert n \rangle \langle n \vert,
\qquad [N , c^{\pm}] = \pm c^{\pm}
      \label{thealgebraAkappas}
      \end{eqnarray}
which easily follow from (\ref{cplus}) and (\ref{cmoins}). It should
be observed that the difference between the operators $c^{\pm}$ and
$b^{\pm}$ manifests itself in (\ref{thealgebraAkappas}) by the
summation from $n=s$ to $n=d(\kappa)$.

\section{Mutually Unbiased Bases}
\subsection{Quantization of the Phase Parameter}
We now examine the consequence of a discretization of the parameter
$\varphi$ in the su(2) case ($\kappa < 0$). By taking ({\em cf}.
\cite{DaoKib2010})
   \begin{eqnarray}
\varphi = - \pi \frac{2j}{2j+1} a  \ \ \Leftrightarrow  \ \ \kappa \varphi = \frac{\pi}{2j+1} a,
\qquad a = 0, 1, \ldots, 2j
   \label{phidiscrete}
   \end{eqnarray}
the state vector $\vert \varphi , \alpha \rangle$ becomes
     \begin{eqnarray}
\vert \varphi , \alpha \rangle \equiv \vert a \alpha \rangle = \frac{1}{\sqrt{2j+1}} \sum_{m = -j}^{j}
q^{(j+m) (j-m+1) a/2 + (j+m) \alpha} \vert j,m \rangle
      \label{state aalpha}
      \end{eqnarray}
The phase operator $E_d$ is of course $\varphi$-dependent. For the
quantized values of $\varphi$ given by (\ref{phidiscrete}),
Equations~(\ref{action1 de Ed}) and (\ref{action2 de Ed}) can be
rewritten as
      \begin{eqnarray}
E_d \vert j , m \rangle = q^{(m-1)a} \vert j , m-1 \rangle  \ \ {\rm for}  \ \ m \not= -j
    \label{action1 de Ed en a}
      \end{eqnarray}
and
      \begin{eqnarray}
E_d \vert j , -j \rangle = q^{ja} \vert j , j \rangle  \ \ {\rm for}  \ \ m = -j
    \label{action2 de Ed en a}
      \end{eqnarray}
The corresponding operator $E_d$ is thus $a$-dependent. However, the
eigenvalues of $E_d$ do not depend on $a$ as shown by
(\ref{eigenvalue Ed}).

\subsection{Connecting the Phase Operator with a Quantization Scheme}
The eigenvector $\vert a \alpha \rangle$ of $E_d$ given by
(\ref{state aalpha}) is a particular case, corresponding to $r=0$,
of the vector
      \begin{eqnarray}
\vert j \alpha ; r a \rangle = \frac{1}{\sqrt{2j+1}} \sum_{m = -j}^{j}
q^{(j+m) (j-m+1) a/2 - jmr + (j+m) \alpha} \vert j,m \rangle
      \label{state jalphara}
      \end{eqnarray}
obtained from a polar decomposition of su(2)
\cite{WRa,CCCC05,IJMP1,IJMP2,AlbKib1}. More precisely
      \begin{eqnarray}
\vert a \alpha \rangle = \vert j \alpha ; 0 a \rangle
      \end{eqnarray}
In quantum information, $\vert a \alpha \rangle$ can represent a
qudit in dimension $d= 2j+1$. The case of a qubit corresponds to $d
= 2$, {\em i.e.}, to an angular momentum $j=1/2$.

The vector $\vert j \alpha ; r a \rangle$ is an eigenvector of the operator
          \begin{eqnarray}
          v_{ra} := {e}^{2 \pi \ii j r} |j , -j \rangle \langle j , j|
                  + \sum_{m = -j}^{j-1} q^{(j-m)a} |j , m+1 \rangle \langle j , m|
          \label{definition of vra}
          \end{eqnarray}
where $r \in \mathbb{R}$ and $a \in \mathbb{Z}/(2j+1)\mathbb{Z}$.
The action of $v_{ra}$ on $|j , m \rangle$ reads
          \begin{eqnarray}
          v_{ra} |j , m \rangle = \delta_{m,j} {e}^{2 \pi \ii j r} |j , -j \rangle
                                + (1 - \delta_{m,j}) q^{(j-m)a} |j , m+1 \rangle
          \end{eqnarray}
and the matrix elements of $v_{ra}$ in the basis $B_{2j+1}$ are
          \begin{eqnarray}
          \langle j , m| v_{ra} |j , m' \rangle = \delta_{m,-j} \delta_{m',j} {e}^{2 \pi \ii j r} |j , -j \rangle
                                                + \delta_{m,m'+1} (1 - \delta_{m',j}) q^{(j-m')a} |j , m+1 \rangle
          \end{eqnarray}
where $m, m' = j, j-1, \ldots, -j$.

As a matter of fact, we have the eigenvalue equation
          \begin{eqnarray}
          v_{ra} \vert j \alpha ; r a \rangle = q^{j(r+a) - \alpha} \vert j \alpha ; r a \rangle,
          \qquad \alpha = 0, 1, \ldots, 2j
          \label{eigenequation for vra}
          \end{eqnarray}
The spectrum of $v_{ra}$ is not degenerate. The vectors $\vert j \alpha ; r a \rangle$
are common eigenvectors of $J^2$ and $v_{ra}$. For fixed $r$ and $a$,   they satisfy
the orthogonality relation
      \begin{eqnarray}
\langle j \alpha ; r a | j \beta ; r a \rangle = \delta_{\alpha,\beta}
      \label{jalphabetara}
      \end{eqnarray}
for $\alpha , \beta = 0, 1, \ldots, 2j$.

The operator $v_{ra}$ is unitary and it commutes with the Casimir
operator $J^2$ of SU(2). The set $\{ J^2 , v_{ra} \}$ is a complete
set of commuting operators that provides an alternative to the
scheme $\{ J^2 , J_z \}$, used in angular momentum theory. In other
words, for fixed $j$, $r$ and $a$, the set
      \begin{eqnarray}
B_{ra} := \{ \vert j \alpha ; r a \rangle : \alpha = 0, 1, \ldots, 2j \}
      \label{basis Bra}
      \end{eqnarray}
constitutes a nonstandard orthonormal basis for the
$(2j+1)$-dimensional irreducible representation of SU(2). The basis
$B_{ra}$ is an alternative to the canonical basis $B_{2j+1}$ defined
in (\ref{canonical basis}). The reader may consult \cite{WRa,CCCC05}
for a study of the $\{ J^2 , v_{ra} \}$ scheme and of its associated
Wigner-Racah algebra.

The $a$-dependent operator $E_d$ and the operator $v_{ra}$ are
closely connected. Indeed, it can be \linebreak checked that
      \begin{eqnarray}
E_d = q^{ja} (v_{0a})^{\dagger} = e^{2 \pi \ii j a / (2j+1)} (v_{0a})^{\dagger}
    \label{connection Ed vra}
      \end{eqnarray}
as can be guessed from (\ref{eigenvalue Ed}) and (\ref{eigenequation for vra}).

\subsection{Introduction of Mutually Unbiased Bases}
The case $r=0$ deserves a special attention. Let us examine the
inner product $\langle a \alpha \vert b \beta \rangle$ of the
vectors $\vert a \beta \rangle$ and $\vert b \beta \rangle$ defined
by (\ref{state aalpha}), in view of its importance in the study of
mutually unbiased bases (MUBs).

For $a = b$, we have
      \begin{eqnarray}
\langle a \alpha \vert a \beta \rangle = \delta_{\alpha,\beta}
      \label{orthonormality}
      \end{eqnarray}
Therefore, for fixed $j$ and $a$ ($2j \in \mathbb{N}$
and $a$ in the ring $\mathbb{Z}/(2j+1)\mathbb{Z}$), the basis
      \begin{eqnarray}
B_{0a} := \{ \vert a \alpha \rangle : \alpha = 0, 1, \ldots, 2j \}
      \label{basis B0a}
      \end{eqnarray}
(a particular case of the basis $B_{ra}$) and the basis $B_{2j+1}$
are interrelated via
      \begin{eqnarray}
\langle j,m \vert a \alpha \rangle = \frac{1}{\sqrt{2j+1}}
q^{(j+m) (j-m+1) a/2 + (j+m) \alpha}  \ \ \Rightarrow  \ \ \vert \langle j,m \vert a \alpha \rangle \vert = \frac{1}{\sqrt{2j+1}}
      \label{jm-aalpha}
      \end{eqnarray}
with $\alpha = 0, 1, \ldots, 2j$ and $m = j, j-1, \ldots, -j$. In
view of (\ref{jm-aalpha}), we see that $B_{0a}$ (and more generally
$B_{ra}$) can be considered as a generalized Fourier transform of
$B_{2j+1}$.

For $a \not= b$, the inner product $\langle a \alpha \vert b \beta
\rangle$ can be expressed in term of the generalized quadratic Gauss
sum defined by (see \cite{Berndt98})
      \begin{eqnarray}
S(u, v, w) := \sum_{k=0}^{|w|-1} e^{\ii \pi (uk^2 + vk) / w}
      \label{Gauss sum}
      \end{eqnarray}
In fact, we have
      \begin{eqnarray}
\langle a \alpha \vert b \beta \rangle = \frac{1}{2j+1} S(u, v, w)
      \label{inner product and Gauss}
      \end{eqnarray}
where
      \begin{eqnarray}
u := a-b, \qquad v := - (a-b)d - 2(\alpha - \beta), \qquad w := d = 2j+1
      \label{quatrevingt}
      \end{eqnarray}
The sum $S(u, v, w)$ can be calculated in the situation where $u$, $v$ and $w$ are integers such that $u$ and $w$
are mutually prime, $u w$ is not zero, and $uw + v$ is even.

Let us now briefly discuss the reason why (\ref{state aalpha}) is of
interest for the determination of MUBs. We recall that two
orthonormal bases of the $d$-dimensional Hilbert space
${\mathbb{C}}^d$ are said to be unbiased if the modulus of the inner
product of any vector of one basis with any vector of the other one
is equal to $1/\sqrt{d}$ \cite{Ivanovic,WoottersFields}. For fixed
$d$, it is known that the number $N_{\scriptscriptstyle MUB}$ of
MUBs is such that $3 \leq N_{\scriptscriptstyle MUB} \leq d+1$ and
that the limit $N_{\scriptscriptstyle MUB}=d+1$ is attained when $d$
is a power of a prime number \cite{Ivanovic,WoottersFields}. Then,
equation (\ref{jm-aalpha}) shows that any basis $B_{0a}$ ($a \in
\mathbb{Z}/(2j+1)\mathbb{Z}$) is unbiased with $B_{2j+1}$ for
arbitrary value of $2j+1$. Furthermore, in the special case where
$2j+1$ is a prime integer, the calculation of $S(u, v, w)$ with
(\ref{quatrevingt}) leads to
      \begin{eqnarray}
\vert \langle a \alpha \vert b \beta \rangle \vert = \frac{1}{\sqrt{2j+1}}
      \label{MUBs condition}
      \end{eqnarray}
for $a \not=b$, $\alpha = 0, 1, \ldots, 2j$ and $\beta = 0, 1,
\ldots, 2j$. Equation (\ref{MUBs condition}) implies that $B_{0a}$
and $B_{0b}$ for $a$ and $b$ in the Galois field $\mathbb{F}_{2j+1}$
are mutually unbiased.

Thus one arrives at the following conclusion. For $2j+1$ prime, the
$2j+1$ bases $B_{0a}$ ($a = 0, 1, \ldots, 2j$) and the basis
$B_{2j+1}$ form a complete set of $d+1 = 2j+2$ MUBs. This result is
in agreement with the one derived in
\cite{IJMP1,IJMP2,AlbKib1,AlbKib2,Kib-quons,CCCC08,JPhysA08,JPhys09,KibDubna}.
It can be extended to the case $r \not= 0$ as follows. For
arbitrarily fixed $r$ and $2j+1$ prime, the $2j+1$ bases $B_{ra}$
($a = 0, 1, \ldots, 2j$) and the basis $B_{2j+1}$ form a complete
set of $d+1 = 2j+2$ MUBs. The parameter $r$ serves to differentiate
various families (or complete sets) of MUBs.

\section{Discrete Fourier Transforms}

We discuss in this section two quadratic versions of the discrete
Fourier transform (DFT), namely, the quantum DFT that connects state
vectors in an Hilbert space and the classical DFT used in signal
analysis.

\subsection{Quantum Quadratic Discrete Fourier Transform}
Equation (\ref{state jalphara}) shows that the vector $\vert j
\alpha ; r a \rangle$ can be considered as a quantum DFT that is
quadratic (in $m$) for $a \not=0$. This transform is nothing but a
quantum ordinary DFT for $r = a = 0$ \cite{VourdasDFT}. For fixed
$j$, $r$ and $a$, the inverse transform is
     \begin{eqnarray}
\vert j,m \rangle = q^{-(j+m) (j-m+1) a/2 + jmr} \frac{1}{\sqrt{2j+1}}
\sum_{\alpha=0}^{2j} q^{-(j+m) \alpha} \vert j \alpha ; r a \rangle
      \label{inverse of state jalphara}
      \end{eqnarray}
Compact relations, more adapted to the Fourier transform formalism,
can be obtained by going back to the change of notation given by
(\ref{identification SU2}) and (\ref{identification dimensions
SU2}). Then, Equations~(\ref{state jalphara}) and (\ref{inverse of
state jalphara}) read
     \begin{eqnarray}
\vert j \alpha ; r a \rangle = q^{(d-1)^2 r / 4}
\frac{1}{\sqrt{d}} \sum_{n = 0}^{d-1}
q^{n(d -n) a/2 + n[\alpha -(d-1)r/2]} \vert n \rangle
      \label{state jalphara en n}
      \end{eqnarray}
and
     \begin{eqnarray}
\vert n \rangle = q^{-n(d -n) a/2 - (d-1)^2 r / 4 + n(d-1)r/2} \frac{1}{\sqrt{d}}
\sum_{\alpha=0}^{d-1} q^{-\alpha n} \vert j \alpha ; r a \rangle
      \label{inverse of state jalphara en n}
      \end{eqnarray}
We shall put
      \begin{eqnarray}
({\bf F_{ra}})_{n \alpha} :=
\frac{1}{\sqrt{d}} q^{n(d -n) a/2 + (d-1)^2 r / 4 + n[\alpha -(d-1)r/2]}
      \label{def Fra}
      \end{eqnarray}
or
      \begin{eqnarray}
({\bf F_{ra}})_{n \alpha} =
\frac{1}{\sqrt{d}} e^{2 \pi \ii f / d}  \ \ {\rm with}  \ \
f := \onefour (d-1)^2 r + \onehalf [ 2 \alpha + d a - (d-1)r ] n - \onehalf a n^2
      \label{def Fra en e}
      \end{eqnarray}
a relation that defines (for fixed $d$, $r$ and $a$) a $d \times d$
matrix ${\bf F_{ra}}$. Let us recall that for a fixed value of $d$
in $\mathbb{N}^*$, both $r$ and $a$ have a fixed value ($r \in
\mathbb{R}$ and $a \in \mathbb{Z}/d\mathbb{Z}$) and $n, \alpha = 0,
1, \ldots, d-1$.

For $d = 2j+1$ arbitrary, we can show that
    \begin{eqnarray}
( ({\bf F_{ra}})^{\dagger} {\bf F_{sb}} )_{\alpha \beta} = \langle j \alpha ; ra \vert j \beta ; sb \rangle
    \label{matrix Fra and inner product}
      \end{eqnarray}
Therefore, in the particular case $r = s$ and $d=p$, where $p$ is prime, we have
    \begin{eqnarray}
\vert ( ({\bf F_{ra}})^{\dagger}
         {\bf F_{rb}} )_{\alpha \beta} \vert =
\vert \langle j \alpha ; ra \vert j \beta ; rb \rangle \vert =
\vert \langle a \alpha \vert b \beta \rangle \vert = \frac{1} {\sqrt{p}}  \ \ {\rm for}  \ \ a \not=b
    \label{cas r et s nuls et d prime}
      \end{eqnarray}
Equation (\ref{cas r et s nuls et d prime}) shall be discussed below
in terms of Hadamard matrices.

\subsection{Quadratic Discrete Fourier Transform}
\subsubsection{6.2.1. Factorization of the Quadratic DFT}
We are now prepared for discussing the transforms (\ref{state
jalphara en n}) and (\ref{inverse of state jalphara en n}) in the
language of classical signal theory. Let us consider the
transformation
    \begin{eqnarray}
x = \{ x_{m    }  \in {\mathbb{C}} : m = 0, 1, \ldots, d-1 \}  \ \ \leftrightarrow  \ \
y = \{ y_{n    }  \in {\mathbb{C}} : n = 0, 1, \ldots, d-1 \}
    \label{transformation x - y}
      \end{eqnarray}
defined by
    \begin{eqnarray}
y_{n} = \sum_{m = 0}^{d-1}           \left( {\bf F_{ra}} \right)_{m n}  x_{m} \ \ \Leftrightarrow \ \
x_{m} = \sum_{n = 0}^{d-1} \overline{\left( {\bf F_{ra}} \right)_{m n}} y_{n}
    \label{transfo yx et xy en Fra}
      \end{eqnarray}
The particular case $r = a = 0$ corresponds to the ordinary DFT. For
$a \not= 0$, the bijective transformation $x \leftrightarrow y$ can
be thought of as a quadratic DFT. The analog of the
Parseval-Plancherel theorem for the ordinary DFT can be expressed in
the following way. The quadratic transformations $x \leftrightarrow
y$ and $x' \leftrightarrow y'$ associated with the same matrix ${\bf
F_{ra}}$, $r \in \mathbb{R}$ and $a \in \mathbb{Z}/d\mathbb{Z}$,
satisfy the conservation rule
    \begin{eqnarray}
\sum_{n = 0}^{d-1} \overline{y_{n}} y'_{n} =
\sum_{m = 0}^{d-1} \overline{x_{m}} x'_{m}
    \label{Parseval-Plancherel}
      \end{eqnarray}
where both sums do not depend on $r$ and $a$.

The matrix ${\bf F_{ra}}$ can be factorized as
    \begin{eqnarray}
{\bf F_{ra}} = {\bf D_{ra}} {\bf F}, \qquad {\bf F} := {\bf F_{00}}
    \label{factorization of Fra}
      \end{eqnarray}
where ${\bf D_{ra}}$ is the $d \times d$ diagonal matrix with the matrix elements
    \begin{eqnarray}
({\bf D_{ra}})_{m n} := q^{m(d -m) a/2 + (d-1)^2 r / 4 - m(d-1)r/2} \delta_{m , n}
    \label{matrix D}
      \end{eqnarray}
For fixed $d$, there are one $d$-multiple infinity of Gaussian
matrices ${\bf D_{ra}}$ (and thus ${\bf F_{ra}}$) distinguished by
$a \in \mathbb{Z}/d\mathbb{Z}$ and $r \in \mathbb{R}$. On the other
hand, ${\bf F}$ is the well-known ordinary DFT matrix. The matrix
${\bf F}$ was the object of a great number of studies. The main
properties of the ordinary DFT matrix ${\bf F}$ are summed up in the
appendix.

\subsubsection{6.2.2. Hadamard Matrices}
The matrix ${\bf F_{ra}}$ defined by (\ref{def Fra}) is unitary. The
modulus of each of its matrix elements is equal to $1/{\sqrt{d}}$.
Thus, ${\bf F_{ra}}$ can be considered as a generalized Hadamard
matrix (we adopt here the normalization of Hadamard matrices
generally used in quantum information and quantum computing)
\cite{AlbKib1,AlbKib2,Kib-quons,CCCC08,JPhysA08,JPhys09}.

In the case where $d$ is a prime number, Equation~(\ref{cas r et s
nuls et d prime}) shows that the matrix $({\bf F_{ra}})^{\dagger}
{\bf F_{rb}}$ is another Hadamard matrix. However, it should be
mentioned that, given two Hadamard matrices ${\bf M}$ and ${\bf N}$,
the product ${\bf M}^{\dagger} {\bf N}$ is not in general a Hadamard
matrix.

\subsubsection{6.2.3. Trace Relations}
The trace of ${\bf F_{ra}}$ reads
    \begin{eqnarray}
{\rm tr} \, {\bf F_{ra}} = e^{\ii \pi (d-1)^2 / (2d)} \frac{1}{\sqrt{d}} S(u, v, w)
    \label{trace of Fra}
      \end{eqnarray}
where $S(u, v, w)$ is given by (\ref{Gauss sum}) with
    \begin{eqnarray}
u := 2-a, \qquad v := d(a-r) + r, \qquad w := d
    \label{x-y-z}
      \end{eqnarray}
Note that the case $a = 2$ deserves a special attention. In this
case, the quadratic character of ${\rm tr} \, {\bf F_{ra}}$
disappears. In addition, if $r = 0$ we get
    \begin{eqnarray}
{\rm tr} \, {\bf F_{02}} = \sqrt{d}
    \label{trace of F02}
      \end{eqnarray}
as can be seen from direct calculation.

\subsubsection{6.2.4. Diagonalization}
It is a simple matter of calculation to prove that
      \begin{eqnarray}
\left( {\bf F_{ra}} \right)^{\dagger} {\bf V}_{ra} {\bf F_{ra}} =
q^{(d-1)(r+a) / 2} \pmatrix{
q^{1}                &      0 &       \ldots &          0    \cr
0                    & q^{2}  &       \ldots &          0    \cr
\vdots               & \vdots &       \ldots &     \vdots    \cr
0                    &      0 &       \ldots & q^{d}         \cr
}
     \label{endomor}
     \end{eqnarray}
where the
matrix
      \begin{eqnarray}
{\bf V_{ra}} :=
\pmatrix{
0                    &    q^a &      0  & \ldots &          0    \cr
0                    &      0 & q^{2a}  & \ldots &          0    \cr
\vdots               & \vdots & \vdots  & \ldots &     \vdots    \cr
0                    &      0 &      0  & \ldots & q^{(d-1)a}    \cr
e^{\ii \pi (d-1) r  }  &      0 &      0  & \ldots &          0    \cr
}
     \label{matrix Vra}
     \end{eqnarray}
represents the linear operator $v_{ra}$ defined by (\ref{definition
of vra}). Therefore, the matrix ${\bf F_{ra}}$ reduces the
endomorphism associated with the operator $v_{ra}$.

Concerning (\ref{endomor}) and (\ref{matrix Vra}), it is important
to note the following conventions. According to the tradition in
quantum mechanics and quantum information, the matrix ${\bf V_{ra}}$
of the operator $v_{ra}$ is set up on the basis $B_{2j+1}$ ordered
from left to right and from top to bottom in the range $\vert j , j
\rangle \equiv \vert d-1 \rangle,
 \vert j , j-1 \rangle \equiv \vert d-2 \rangle, \ldots,
 \vert j , -j  \rangle \equiv \vert 0   \rangle$. For the sake of compatibility,
we adopt a similar convention for the other matrices under
consideration. Thus, the lines and columns of ${\bf F_{ra}}$ are
arranged in the order $d-1, d-2, \ldots, 0$.

\subsubsection{6.2.5. Link with the Cyclic Group}
There exists an interesting connection between the matrix ${\bf X}$ and the cyclic group
$C_d$ \cite{IJMP1,IJMP2,AlbKib1}. Let us call $R$ a rotation of $2 \pi / d$ around
an arbitrary axis, the generator of $C_d$. Then, the application
       \begin{eqnarray}
C_d \to \{ {\bf X}^n : n = 0, 1, \ldots, d-1 \} \ : \ R \mapsto {\bf X}
       \end{eqnarray}
defines a $d$-dimensional matrix representation of $C_d$. This
representation is the regular representation of $C_d$. Thus, the
reduction of the representation $\{ {\bf X}^n : n = 0, 1, \ldots,
d-1 \}$ contains once and only once each (one-dimensional)
irreducible representation of $C_d$.

\subsubsection{6.2.6. Decomposition}
The matrix ${\bf V_{ra}}$ can be decomposed as
     \begin{eqnarray}
{\bf V_{ra}} = {\bf P_r} {\bf X} {\bf Z}^a
     \label{Vra en X et X}
     \end{eqnarray}
where
       \begin{eqnarray}
{\bf P_r} :=
\pmatrix{
1                    &      0 &      0    & \ldots &       0                 \cr
0                    &      1 &      0    & \ldots &       0                 \cr
0                    &      0 &      1    & \ldots &       0                 \cr
\vdots               & \vdots & \vdots    & \ldots &  \vdots                 \cr
0                    &      0 &      0    & \ldots &       e^{\ii \pi (d-1) r} \cr
}
        \label{definition of P}
        \end{eqnarray}
and
       \begin{eqnarray}
{\bf X} :=
\pmatrix{
0                    &      1 &      0  & \ldots &       0 \cr
0                    &      0 &      1  & \ldots &       0 \cr
\vdots               & \vdots & \vdots  & \ldots &  \vdots \cr
0                    &      0 &      0  & \ldots &       1 \cr
1                    &      0 &      0  & \ldots &       0 \cr
}, \qquad
{\bf Z} :=
\pmatrix{
1                    &      0 &      0    & \ldots &       0       \cr
0                    &      q &      0    & \ldots &       0       \cr
0                    &      0 &      q^2  & \ldots &       0       \cr
\vdots               & \vdots & \vdots    & \ldots &  \vdots       \cr
0                    &      0 &      0    & \ldots &       q^{d-1} \cr
}
        \label{definition of X and Z}
        \end{eqnarray}
The matrices ${\bf P_r}$, ${\bf X}$ and ${\bf Z}$ (and thus ${\bf V_{ra}}$) are unitary. They satisfy
     \begin{eqnarray}
{\bf V_{ra}} {\bf Z} &=& q      {\bf Z} {\bf V_{ra}} \label{qcom 1} \\
{\bf V_{ra}} {\bf X} &=& q^{-a} {\bf X} {\bf V_{ra}} \label{qcom 2}
     \end{eqnarray}
Equation (\ref{qcom 1}) can be iterated to give the useful relation
     \begin{eqnarray}
({\bf V_{ra}})^m {\bf Z}^n = q^{mn}
{\bf Z}^n ({\bf V_{ra}})^m
     \label{VmZn}
     \end{eqnarray}
where $m,n \in \mathbb{Z}/d\mathbb{Z}$. Furthermore, we have the quasi-nilpotency relations
     \begin{eqnarray}
e^{-\ii \pi (d-1) r} ({\bf V_{r0}})^d = {\bf Z}^d = {\bf I_d}
     \label{nilpotency}
     \end{eqnarray}
(the relations (\ref{nilpotency}) are true nilpotency relations when $r = 0$). More
generally, we obtain
     \begin{eqnarray}
\forall n \in \mathbb{Z}/d\mathbb{Z} : ({\bf V_{ra}})^n = q^{-n(n-1)a/2} ({\bf V_{r0}})^n {\bf Z}^{an}
 \ \ \Rightarrow  \ \ ({\bf V_{ra}})^d = e^{\ii \pi (d-1) (r+a)} {\bf I_d}
     \end{eqnarray}
in agreement with the obtained eigenvalues for ${\bf V_{ra}}$ (see Equation~(\ref{eigenequation for vra})).

\subsubsection{6.2.7. Weyl Pairs}
For $r = a = 0$, Equations~(\ref{qcom 1}) and (\ref{nilpotency})
show that the unitary matrices ${\bf X}$ and ${\bf Z}$ satisfy the
$q$-commutation relation
     \begin{eqnarray}
{\bf X} {\bf Z} = q {\bf Z} {\bf X}
     \label{XZ equal qZX}
     \end{eqnarray}
and the nilpotency relations
     \begin{eqnarray}
     {\bf X}^d =
     {\bf Z}^d = {\bf I_d}
     \label{nilpotency pour X et Z}
     \end{eqnarray}
Therefore, ${\bf X}$ and ${\bf Z}$ constitute a Weyl pair (${\bf X}
, {\bf Z}$). Note that the Weyl pair (${\bf X} , {\bf Z}$) can be
defined from the matrix ${\bf V_{ra}}$ only since
    \begin{eqnarray}
{\bf X} = {\bf V_{00}}, \qquad {\bf Z} = \left( {\bf V_{00}} \right)^{\dagger} {\bf V_{01}}
    \end{eqnarray}
which emphasize the important role played by the matrix ${\bf
V_{ra}}$. Note also that according to (\ref{endomor}), we have
      \begin{eqnarray}
      {\bf F}^{\dagger} {\bf X} {\bf F} = q {\bf Z}
      \label{endomor pour X et Z}
      \end{eqnarray}
that proves that ${\bf X}$ and ${\bf Z}$ are related by the DFT matrix.

Weyl pairs were introduced at the beginning of quantum mechanics
\cite{Weyl} and used for building operator unitary bases
\cite{Schwinger}. The pair (${\bf X} , {\bf Z}$) plays an important
role in quantum information and quantum computing. In these fields,
the linear operators corresponding to ${\bf X}$ and ${\bf Z}$ are
known as flip or shift and clock operators, respectively. For $d$
arbitrary, they are at the root of the Pauli group, a finite
subgroup of order $d^3$ of the group U($d$) for $d$ even and SU($d$)
for $d$ odd \cite{JPhysA08,JPhys09}. The Pauli group is of
considerable importance for describing quantum errors and quantum
fault tolerance in quantum computation (see \cite{G1,G2,G3,JCTN} and
references therein for recent geometrical approaches to the Pauli
group). The Weyl pair (${\bf X} , {\bf Z}$) turns out to be an
integrity basis for generating the set $\{ {\bf X}^a {\bf Z}^b : a,b
\in \mathbb{Z}/d\mathbb{Z} \}$. The latter set constitutes a basis
for the Lie algebra of the unitary group U($d$) with respect to the
commutator law. This set consists of $d^2$ generalized Pauli
matrices in $d$ dimensions \cite{JPhysA08,JPhys09}. In this respect,
note that for $d=2$ we have
   \begin{eqnarray}
{\bf X}      =     {\bf \sigma_x}, \qquad
{\bf Z}      =     {\bf \sigma_z}, \qquad
{\bf XZ}     = - i {\bf \sigma_y}, \qquad
{\bf X^0Z^0} =     {\bf \sigma_0}
   \end{eqnarray}
in terms of the ordinary Pauli matrices ${\bf \sigma_0} = {\bf I_2}$, ${\bf \sigma_x}$, ${\bf \sigma_y}$, and $\bf \sigma_z$.

\subsubsection{6.2.8. Link with a Lie Algebra}
Equation (\ref{VmZn}) can be particularized to give
   \begin{eqnarray}
{\bf X}^m {\bf Z}^n = q^{mn}
{\bf Z}^n {\bf X}^m,
\qquad  (m,n) \in \mathbb{N}^2
   \end{eqnarray}
Let us define the operator
   \begin{eqnarray}
T_{(n_1,n_2)} := q^{ {1 \over 2} n_1n_2 } Z^{n_1} X^{n_2}, \qquad
                 (n_1,n_2) \in \mathbb{N}^2
   \end{eqnarray}
It is convenient to use the abbreviation
   \begin{eqnarray}
(n_1, n_2) \equiv n  \ \ \Rightarrow  \ \ T_{n} \equiv T_{(n_1,n_2)}
   \label{92}
   \end{eqnarray}
The product $T_nT_m$ is easily obtained to be
   \begin{eqnarray}
T_m T_n = q^{ - {1 \over 2} m \times n } T_{m + n}
   \label{93}
   \end{eqnarray}
where
   \begin{eqnarray}
m \times n := m_1 n_2 - m_2 n_1, \qquad m + n = (m_1 + n_1, m_2 + n_2)
   \label{94}
   \end{eqnarray}
The commutator $[T_m,T_n]$,
   \begin{eqnarray}
[T_m,T_n] = - 2 {\rm i} \sin \left( { \pi \over k } m \times n \right) T_{m + n}
   \label{95}
   \end{eqnarray}
follows at once from (\ref{93}). The operators $T_m$ can be thus
formally viewed as the generators of the infinite-dimensional Lie
algebra $W_{\infty}$ (or sine algebra) investigated in
\cite{FFZ,DHK}.

\section{Closing Remarks}

We used the representation theory of the symmetry groups SU(2) and
SU(1,1) to describe the generalized oscillator algebra ${\cal
A}_\kappa$ and the two phase operators $E_d$ and $E_{\infty}$
introduced in \cite{DaoKib2010}. The phase eigenstates of $E_d$ and
$E_{\infty}$ were thus understood in terms of finite-dimensional and
infinite-dimensional representations of SU(2) and SU(1,1),
respectively. In the case of those representations of SU(2) for
which the dimension is a prime integer, our approach led us to
derive MUBs as eigenbases of the phase operator $E_d$ (with $d$
prime), opening a way for further results on unitary phase operators
associated with Lie groups.

The unitary phase operator $E_d$ defined via
      \begin{eqnarray}
J_- = E_d \sqrt{J_+ J_-}  \ \ \Leftrightarrow  \ \ J_- = \sqrt{J_+ J_-} (E_d)^{\dagger}
    \label{Jm et Jp en polaire}
      \end{eqnarray}
leads to a polar decomposition of the algebra su(2) in the scheme
$\{ J^2 , E_d \}$, which is an alternative to the familiar
quantization scheme $\{ J^2 , J_3 \}$ of angular momentum theory.
The $\{ J^2 , E_d \}$ scheme and the $\{ J^2 , v_{ra} \}$ scheme of
\cite{WRa,CCCC05,IJMP1,IJMP2,AlbKib1,AlbKib2,Kib-quons,CCCC08,JPhysA08,JPhys09,KibDubna}
are related by (\ref{connection Ed vra}). In the case of the
noncompact Lie algebra su(1,1), the phase operator $E_{\infty}$ is
non-unitary and given by
      \begin{eqnarray}
K_- = E_{\infty} \sqrt{K_+ K_-}  \ \ \Leftrightarrow  \ \ K_+ = \sqrt{K_+ K_-} (E_{\infty})^{\dagger}
    \label{Km et Kp en polaire}
      \end{eqnarray}
Although this does not correspond to a true polar decomposition
(because $E_{\infty}$ is not unitary), it yields a scheme $\{ K^2 ,
E_{\infty} \}$, which is an alternative to the canonical scheme $\{
K^2 , K_3 \}$ developed for su(1,1) by Bargmann and most of other
authors. We hope to further study this new scheme from the point of
view of the representation theory and the Wigner-Racah algebra of
SU(1,1).

As far as the applications of the new SU(2) and SU(1,1) phase states
derived in Section 3 are concerned, let us mention that, besides the
two applications (to mutually unbiased bases in section 5 and to
discrete Fourier transform in Section 6) discussed in our paper, we
can mention other potential applications. Our phase states can be
useful for various dynamical systems (e.g., the Morse system for the
SU(2) states as well as the P\" oschl-Teller system and the
repulsive oscillator system for the SU(1,1) states). We can also
mention a possible application of the quadratic discrete Fourier
transform to discrete linear canonical transforms and to Hadamard
matrices in connection with the production of geometric optics
setups. Some of these further potential applications are presently
under consideration.


\section*{Acknowledgements}
One of the authors (MRK) thanks the {\em Instituto de Matem\'aticas}
and the {\em Instituto de Ciencias F\'{\i}sicas} of the {\em
Universidad National Aut\'onoma de M\'exico} (UNAM) for financial
support and the kind hospitality extended to him during his stay at
the UNAM in Cuernavaca. The authors acknowledge the support of the
{\em \'Optica Matem\'atica} projects (DGAPA-UNAM IN-105008 and
SEP-CONACYT 79899).

\bibliographystyle{mdpi}
\makeatletter
\renewcommand\@biblabel[1]{#1. }
\makeatother

\begin{thebibliography}{100}


\bibitem{Fernandez-1}
Fern\'andez, C.D.J.; Nieto, L.M.; Rosas-Ortiz, O.
Distorted Heisenberg algebra and coherent states for isospectral  oscillator Hamiltonians.
{\em J. Phys. A: Math. Gen.} {\bf 1995}, {\em 28}, 2693--2708.


\bibitem{Fernandez-2}
Fern\'andez, C.D.J.; Hussin, V.
Higher-order SUSY, linearized nonlinear Heisenberg algebras and coherent states.
{\em J. Phys. A: Math. Gen.} {\bf 1999}, {\em 32}, 3603--3619.


\bibitem{Fernandez-3}
Carballo, J.M.; Fern\'andez, C.D.J.; Negro, J.; Nieto, L.M.
Polynomial Heisenberg algebras.
{\em J. Phys. A: Math. Gen.} {\bf 2004}, {\em 37}, 10349--10362.


\bibitem{Plyu96}
Plyushchay, M.S.
Deformed Heisenberg algebra, fractional spin
fields, and supersymmetry without fermions.
{\em Ann. Phys. NY} {\bf 1996}, {\em 245}, 339--360.


\bibitem{Plyu97}
Plyushchay, M.S. Deformed Heisenberg algebra with reflection. {\em Nucl. Phys. B} {\bf 1997}, {\em 491}, 619--634.


\bibitem{Plyu00}
Plyushchay, M.
Hidden nonlinear supersymmetries in pure parabosonic systems.
{\em Int. J. Mod. Phys. A} {\bf 2000}, {\em 15}, 3679--3698.


\bibitem{Plyu10}
Horvathy, P.A.; Plyushchay, M.S.; Valenzuela, M.
Bosons, fermions and anyons in the plane, and supersymmetry.
{\em Ann. Phys. NY}, in press.


\bibitem{Ques1}
Quesne, C; Vansteenkiste, N.
C-lambda-extended harmonic oscillator and (para)supersymmetric quantum mechanics.
{\em Phys. Lett. A} {\bf 1998}, {\em 240}, 21--28.


\bibitem{Ques2}
Quesne, C.
Spectrum generating algebra of the C-lambda-extended oscillator and multiphoton coherent states.
{\em Phys. Lett. A} {\bf 2000}, {\em 272}, 313--325.


\bibitem{Ques3}
Quesne, C.; Vansteenkiste N.
C-lambda-extended oscillator algebras and some of their deformations and applications to quantum mechanics.
{\em Int. J. Theor. Phys.} {\bf 2000}, {\em 39}, 1175--1215.


\bibitem{Ques4}
Quesne, C.
Fractional supersymmetric quantum mechanics, topological
invariants and generalized deformed oscillator algebras.
{\em Mod. Phys. Lett. A} {\bf 2003}, {\em 18}, 515--525.


\bibitem{Kib1}
Daoud, M.; Kibler, M.
A fractional supersymmetric oscillator and its coherent states. In
Proceedings of the Sixth International Wigner Symposium, Istanbul, Turkey, 16--22 August 1999; Engin, A., Ed.;
Bogazici University Press: Istanbul, Turkey, 2002; pp. 125--139.


\bibitem{Kib2}
Daoud, M.; Kibler, M.R.
On fractional supersymmetric quantum mechanics: The fractional supersymmetric oscillator.
In {\em Symmetry and Structural Properties of Condensed Matter};
Lulek, T., Lulek, B., Wal, A., Eds.; World Scientific: Singapore,
2001; pp. 408--421.


\bibitem{Kib3}
Daoud, M.; Kibler, M.
Fractional supersymmetric quantum mechanics as a set of
replicas of ordinary supersymmetric quantum mechanics.
{\em Phys. Lett. A} {\bf 2004}, {\em 321}, 147--151.


\bibitem{Kib4}
Daoud, M.; Kibler, M.R.
Fractional supersymmetry and hierarchy of shape invariant potentials.
{\em J. Math. Phys.} {\bf 2006}, {\em 47}, 122108.


\bibitem{DaoKib2010}
Daoud, M.; Kibler, M.R.
Phase operators, temporally stable phase states, mutually
unbiased bases and exactly solvable quantum systems.
{\em J. Phys. A: Math. Theor.} {\bf 2010}, {\em 43}, 115303.


\bibitem{Pegg}
Pegg, D.T.; Barnett, S.M.
Phase properties of the quantized single-mode electromagnetic-field.
{\em Phys. Rev. A} {\bf 1989}, {\em 39}, 1665--1675.


\bibitem{Roy1}
Roy, P.; Roy, B.
Remarks on the construction of a Hermitian phase operator.
{\em Quantum Semiclass. Opt.} {\bf  1997}, {\em 9}, L37--L44.


\bibitem{Roy2}
Roy, B.; Roy, P.
Coherent states, even and odd coherent states in a
finite-dimensional Hilbert space and their properties.
{\em J. Phys. A: Math. Gen.} {\bf 1998}, {\em 31}, 1307--1317.


\bibitem{Perelomov86}
Perelomov, A.M.
{\em Generalized Coherent States and Their Applications};
Springer: Berlin, Germany, 1986.


\bibitem{Gazeau}
Gazeau, J.-P.
{\em Coherent States in Quantum Physics};
Wiley-VCH: Berlin, Germany, 2009.


\bibitem{WRa}
Kibler, M.R.
On the Wigner-Racah algebra of the group $SU_2$ in a non-standard basis.
In {\em Symmetry and Structural Properties of Condensed Matter};
Lulek, T., Lulek, B., Wal, A., Eds.; World Scientific: Singapore, 1999; pp. 222--233.


\bibitem{CCCC05}
Kibler, M.R.
Representation theory and Wigner-Racah algebra of the group SU(2) in a noncanonical basis.
{\em Collect. Czech. Chem. Commun.} {\bf 2005}, {\em 70}, 771--796.


\bibitem{IJMP1}
Kibler, M.R.
Angular momentum and mutually unbiased bases.
{\em Int. J. Mod. Phys. B} {\bf 2006}, {\em 20}, 1792--1801.


\bibitem{IJMP2}
Kibler, M.R.; Planat M.
A SU(2) recipe for mutually unbiased bases.
{\em Int. J. Mod. Phys. B} {\bf 2006}, {\em 20}, 1802--1807.


\bibitem{AlbKib1}
Albouy, O.; Kibler, M.R.
SU(2) nonstandard bases: Case of mutually unbiased bases.
{\em SIGMA} {\bf 2007}, {\em 3}, 076.


\bibitem{AlbKib2}
Albouy, O.; Kibler, M.R.
A unified approach to SIC-POVMs and MUBs.
{\em J. Russ. Laser Res.} {\bf 2007}, {\em 28}, 429--438.


\bibitem{Kib-quons}
Kibler, M.R.
Miscellaneous applications of quons.
{\em SIGMA} {\bf 2007}, {\em 3}, 092.


\bibitem{CCCC08}
Kibler, M.
Generalized spin bases for quantum chemistry and quantum information.
{\em Collect. Czech. Chem. Commun.} {\bf 2008}, {\em 73}, 1281--1298.


\bibitem{JPhysA08}
Kibler, M.R.
Variations on a theme of Heisenberg, Pauli and Weyl.
{\em J. Phys. A: Math. Theor.} {\bf 2008}, {\em 41}, 375302.


\bibitem{JPhys09}
Kibler, M.R.
An angular momentum approach to quadratic Fourier transform, Hadamard matrices,
Gauss sums, mutually unbiased bases, unitary group and Pauli group.
{\em J. Phys. A: Math. Theor.} {\bf 2009}, {\em 42}, 353001.


\bibitem{KibDubna}
Kibler, M.R.
Bases for qudits from a nonstandard approach to SU(2).
{\em Phys. Atom. Nucl.}, in press.


\bibitem{VouBM96}
Vourdas, A.; Brif, C.; Mann, A.
Factorization of analytic representations in the unit disc
and number-phase statistics of a quantum harmonic oscillator.
{\em J.\ Phys.\ A Math.\ Gen.} {\bf 1996}, {\em 29}, 5887--5898.


\bibitem{Berndt98}
Berndt, B.C.; Evans, R.J.; Williams, K.S.
{\em Gauss and Jacobi Sums}; Wiley: New York, NY, USA, 1998.


\bibitem{Ivanovic}
Ivanovi\'c, I.D.
Geometrical description of quantum state determination.
{\em J. Phys. A: Math. Gen.} {\bf 1981}, {\em 14}, 3241--3245.


\bibitem{WoottersFields}
Wootters, W.K.; Fields, B.D.
Optimal state-determination by mutually unbiased measurements.
{\em Ann. Phys. NY} {\bf 1989}, {\em 191}, 363--381.


\bibitem{VourdasDFT}
Vourdas, A.
Quantum systems with finite Hilbert space.
{\em Rep. Prog. Phys.} {\bf 2004}, {\em 67}, 267--320.


\bibitem{Weyl}
Weyl, H.
{\em The Theory of Groups and Quantum Mechanics};
Dover Publications: New York, NY, USA, 1931.


\bibitem{Schwinger}
Schwinger, J.
Unitary operator bases.
{\em Proc. Nat. Acad. Sci. USA} {\bf 1960}, {\em 46}, 570--579.


\bibitem{G1}
Havl\'{\i}\v{c}ek, H.; Saniga, M.
Projective ring line on an arbitrary single qudit.
{\em J. Phys. A: Math. Theor.} {\bf 2008}, {\em 41}, 015302.


\bibitem{G2}
Planat, M.; Baboin, A.-C.; Saniga, M.
Multi-line geometry of qubit-qutrit and higher-order Pauli operators.
{\em Int. J. Theor. Phys.} {\bf 2008}, {\em 47}, 1127--1135.


\bibitem{G3}
Albouy, O.
The isotropic line of ${\mathbb{Z}_d}^2$.
{\em J. Phys. A: Math. Theor.} {\bf 2009}, {\em 42}, 072001.


\bibitem{JCTN}
Planat, M.; Kibler, M.
Unitary reflection groups for quantum fault tolerance.
{\em J. Comput. Theor. Nanosci.} {\bf 2010}, {\em 7}, 1--12.


\bibitem{FFZ}
Fairlie, D.B.; Fletcher, P.; Zachos, C.K.
Infinite-dimensional algebras and a trigonometric
basis for the classical Lie-algebras.
{\em J. Math. Phys.} {\bf 1990}, {\em 31}, 1088--1094.


\bibitem{DHK}
Daoud, M.; Hassouni, Y.; Kibler, M.
The k-fermions as objects interpolating between fermions and bosons.
In {\em Symmetries in Science X};
Gruber, B., Ramek, M., Eds.; Plenum Press: New York, NY, USA, 1998; pp. 63--77.


\bibitem{KBW-GK-I}
Wolf, K.B.; Kr\"otzsch, G.
Geometry and dynamics in the fractional discrete Fourier transform.
{\em J.\ Opt.\ Soc.\ Am.\ A} {\bf 2007}, {\em 24}, 651--658.


\bibitem{Ozaktas-book}
Ozaktas, H.M.; Zalevsky, Z.; Kutay M.A.
{\em Fractional Fourier Transform with Applications
in Optics and Signal Processing}; Wiley: Chichester, UK, 2001.


\bibitem{Condon}
Condon, E.U.
Immersion of the Fourier transform in a continuous group of functional transformations.
{\em Proc.\ Nat.\ Acad.\ Sci.\ USA} {\bf 1937}, {\em 23}, 158--164.


\bibitem{Collins}
Collins, S.A., Jr.
Lens-system diffraction integral written in terms of matrix optics.
{\em J.\ Opt.\ Soc.\ Am.} {\bf 1970}, {\em 60}, 1168--1177.


\bibitem{Moshinsky-Quesne}
Moshinsky, M.; Quesne, C.
Oscillator systems. In Proceedings of the 15th Solvay Conference in
Physics, 1970; Gordon and Breach: New York, NY, USA, 1974.


\bibitem{Pei1}
  Pei, S.-C.; Yeh, M.-H.
  Improved discrete fractional transform.
  {\em Opt.\ Lett.} {\bf 1997}, {\em 22}, 1047--1049.


\bibitem{Pei2}
  Pei, S.-C.; Tseng, C.-C.
  Discrete fractional Fourier transform based on orthogonal projections.
  {\em IEEE Trans.\ Signal Process.} {\bf 1999}, {\em 47}, 1335--1348.


\bibitem{Harper}
  Barker, L.; Candan, \c C.; Hakio\u glu, T.; Kutay, M.A.; Ozaktas, H.M.
  The discrete harmonic oscillator, Harper's equation, and the discrete fractional Fourier transform.
    {\em J.\ Phys.\ A Math. Gen.} {\bf 2000}, {\em 33}, 2209--2222.


\bibitem{Healy-Sheridan-2010}
Healy, J.J.; Sheridan, J.T.
Fast linear canonical transforms.
{\em J.\ Opt.\ Soc.\ Am.\ A} {\bf 2010}, {\em 27}, 21--30.


\bibitem{Namias}
  Namias, V.
  The fractional order Fourier transform and its application to quantum mechanics.
    {\em J.\ Inst.\ Math.\ Appl.} {\bf 1980}, {\em 25}, 241--265.


\bibitem{Mehta}
  Mehta, M.L.
  Eigenvalues and eigenvectors of the finite Fourier transform.
  {\em J.\ Math.\ Phys.} {\bf 1987}, {\em 28}, 781--785.


\bibitem{Ruzzi}
  Ruzzi, M.
  Jacobi $\vartheta$-functions and discrete Fourier transforms.
  {\em J.\ Math.\ Phys.} {\bf 2006}, {\em 47}, 063507.


\bibitem{q-Fourier}
  Mu\~noz, C.A.; Rueda-Paz, J.; Wolf, K.B.
    Fractional discrete $q$-Fourier transforms.
    {\em J.\ Phys.\ A Math.\ Theor.} {\bf 2009}, {\em 42}, 355212.


\end{thebibliography}

\section*{Appendix: Properties of the Ordinary DFT Matrix}
The ordinary DFT---also called the {\em finite} Fourier transform---is the linear transformation of the
complex $d$-dimensional Hilbert space ${\mathbb{C}}^d$ onto itself, that is represented by the matrix ${\bf F}$
whose elements are given by
      \begin{eqnarray}
({\bf F})_{m n} :=
\frac{1}{\sqrt{d}} q^{m n} = \frac{1}{\sqrt{d}} e^{2 \pi \ii mn / d}
      \label{def F}
      \end{eqnarray}
with $m, n = 0, 1, \ldots, d-1$. The elements $({\bf F})_{m n}$ are periodic in $m$ and $n$ modulo $d$ (so that ${\bf F}$ can
be stitched into a torus), but we shall consider the fundamental
interval to be $0 \le m,n \le d-1$. \linebreak Equation (\ref{def F}) follows from (\ref{def Fra}) and (\ref{factorization of Fra}). The
matrix ${\bf F}$ corresponds to the transformation (\ref{transformation x - y}) with
    \begin{eqnarray}
y_{n} = \frac{1}{\sqrt{d}} \sum_{m = 0}^{d-1} q^{mn}  x_{m}  \ \ \Leftrightarrow  \ \
x_{m} = \frac{1}{\sqrt{d}} \sum_{n = 0}^{d-1} q^{-nm} y_{n}
    \label{transfo yx et xy en F}
      \end{eqnarray}
Note that in the physics literature it is more common to find the definition (\ref{def F})
with a minus sign in the exponent; of course, the results obtained with the two conventions
are equivalent.


The Fourier matrix ${\bf F}$ has several well-known properties. It is symmetric and unitary. In addition \linebreak it satisfies
    \begin{eqnarray}
{\bf F}^4 = {\bf I}_d
    \label{F puissance 4}
      \end{eqnarray}
Because ${\bf F}$ is unitary, its eigenvalues must be on the unit circle $S^1$,
and since it is a fourth root of unity, so are its eigenvalues, to be denoted by
    \begin{eqnarray}
\varphi_k := {\ii}^k = e^{\ii \pi k / 2} \in \{ 1,\ii,-1,-\ii \}
        \label{Fou-eigenvalues}
      \end{eqnarray}
for $k = 0, 1, 2, 3$. This divides the space ${\mathbb{C}}^d$ into four
Fourier-invariant, mutually orthogonal subspaces whose dimensions
$N_{\varphi_k}$ exhibit the modulo-4 multiplicities of the eigenvalues
$\varphi_k$. Of course, we have
    \begin{eqnarray}
d = \sum_{k=0}^3 N_{\varphi_k}, \qquad
\hbox{tr} \, {\bf F} = \sum_{k=0}^3  \varphi_k  N_{\varphi_k}, \qquad
\det {\bf F} = \prod_{k=0}^3 (\varphi_k)^{N_{\varphi_k}}
    \label{properties of F}
      \end{eqnarray}
For $d = 4J + k$ with $k = 0, 1, 2, 3$ and $J \in \mathbb{N}$, the multiplicities, traces and
determinants of the submatrices of ${\bf F}$ associated with each eigenvalue
are given by:
\be
    \begin{array}{lccccrc}
     \hbox{dimension} & &\hbox to0pt{\hss \qquad multiplicities
     $N_{\varphi_k}$\hss} & & &  \\
  {\!\!\!}d{=}&\varphi_0{=}1 & \varphi_1{=}\ii & \varphi_2{=}-1
            & \varphi_3{=}-\ii & \hbox{tr}\,\FF\,{}& \det\FF\\[3pt]
        4J & J{+}1     & J &   J   &   J{-}1  &1+\ii  & \ii(-1)^J \\
        4J{+}1 & J{+}1 &   J   &   J   &   J  &1 &  (-1)^J\\
        4J{+}2 & J{+}1 &   J   & J{+}1 &   J  &0 &  -(-1)^J\\
        4J{+}3 & J{+}1 &   J{+}1   & J{+}1 & J&  \ii & -\ii(-1)^J
                \end{array}             \label{Fou-multiplciity}
\ee
(see for example \cite{KBW-GK-I} noting that the DFT matrix there is the complex conjugate of the DFT matrix here).

Since $N_{\varphi_k} \approx d / 4$, there is wide freedom in choosing
eigenvector bases within each eigenspace. Finding a ``good''
eigenbasis is of interest to define fractional powers of the DFT
matrices, which constitute the abelian group of elements $\{\FF^\nu\}$,
for real $\nu$ modulo 4 \cite{Ozaktas-book}, that would
contract, for $d\to\infty$, to the fractional
Fourier {\em integral} transform. The fractionalization of the Fourier integral transform was
defined in 1937 by Condon \cite{Condon} at the suggestion of von Neumann, rediscovered in other contexts
\cite{Collins,Moshinsky-Quesne}, and is currently of importance for signal analysis and image processing
through the fast Fourier transform algorithm \cite{Pei1,Pei2,Harper,Healy-Sheridan-2010}. The
integral kernel of the fractional Fourier integral transform can be expressed as a bilinear
generating function for Hermite-Gauss functions \cite{Namias},
\be
    \Psi_n(x):= \frac{1}{\sqrt{2^n\,n!\,\surd\pi}} e^{- x^2 / 2}
                H_n(x)     \label{Hermite-Gauss}
\ee
where $H_n(x)$ are the Hermite polynomials of degree $n \in \mathbb{N}$ in $x$, which are the eigenfunctions of
the Fourier integral transform $\cal F$,
\be
    \widetilde\Psi_n(x)=({\cal F}\,\Psi_n)(x)
        =\frac1{\surd2\pi} \int_{-\infty}^{+\infty} \dd x'\,
            e^{\ii x x'}\,\Psi_n(x') = {\ii}^n\Psi_n(x)
                            \label{Eigen-FHGauss}
\ee
The integral kernel of the fractional Fourier integral transform \cite{Ozaktas-book} is then
obtained as
             \begin{eqnarray}
     F^\nu(x,x')&:=&\frac1{\sqrt{2\pi\sin\onehalf\pi\nu}}
        \exp \Big( \ii \frac{xx'-\onehalf(x^2{+}x^{\prime2})
                \cos\onehalf\pi\nu}{\sin\onehalf\pi\nu}\Big)
                            \label{FracF1}\\
            &=&\sum_{n=0}^\infty \Psi_n(x)\,
                    e^{\ii \pi \nu / 2}\,\Psi_n(x')
                            \label{FracF2}
             \end{eqnarray}
for $0< \nu <2$, with the limits
    \begin{eqnarray}
F^0(x,x') = \delta(x-x'), \qquad
F^2(x,x') = \delta(x+x')
    \end{eqnarray}
These kernels are unitary,
    \begin{eqnarray}
F^{-\nu}(x,x') = \overline{F^{\nu}(x',x)}
    \end{eqnarray}
and form a one-parameter group
    \begin{eqnarray}
\int_{-\infty}^{+\infty} \dd x'\, F^{\nu_1}(x,x')\,F^{\nu_2}(x',x'')
=F^{\nu_1+\nu_2}(x,x'')
    \end{eqnarray}
with $\nu$ modulo 4.

To fractionalize the DFT matrix \FF, one will be naturally
interested finding $d$-point functions $\Phi^{(d)}_n(m)$ that are
``good'' discrete counterparts for the Hermite-Gauss functions
$\Psi_n(x)$ in (\ref{Hermite-Gauss}); in particular that they be {\em
analytic\/} and {\em periodic\/} functions of $m$. Mehta \cite{Mehta}
has proposed the following functions:
\be
    \Phi^{(d)}_n(m):=\sum_{\ell=-\infty}^\infty
        \exp\Big( {-\frac\pi{d}}(\ell d+m)^2\Big)
            H_n\Big(\sqrt{\frac{2\pi}d}(\ell d+m)\Big)
                    \label{Mehta-func}
\ee
that we call {\em Mehta\/} functions. These have the desired
properties and
\be
    \FF\,{\bf\Phi}^{(d)}_n= {\ii}^n\,{\bf\Phi}^{(d)}_n
                    \label{eigen-Mehta}
\ee
for $n \in \mathbb{N}$ and where ${\bf \Phi}^{(d)}_n$ is the column vector
of components $\Phi^{(d)}_n(m)$. Of course, there cannot be more than $d$
linearly independent vectors in ${\mathbb{C}}^d$, so we may take the
subset $\{ {\bf\Phi}^{(d)}_n : n = 0, 1, \ldots, d-1 \}$. {\em Prima facie},
it is not clear whether this subset is
linearly independent and orthogonal, or not -- Mehta \cite{Mehta}
left unresolved their orthogonality, which was lately described
thoroughly by Ruzzi \cite{Ruzzi}. The departure from strict orthogonality
of the vectors of the Mehta basis   was investigated in \cite{q-Fourier}; the departure is small for low
values of $n$ and gradually worsens up to $d-1$.

Indeed, there is wide freedom in choosing bases for ${\mathbb{C}}^d$ when
the sole requirement is that they be eigenbases of \FF, satisfying
(\ref{eigen-Mehta}). Labelling these eigenvectors by their four Fourier
eigenvalues $\varphi_k$, and within each of these eigenspaces
${{\mathbb{C}}}^{N_{\varphi_k}}$ by $j = 0, 1, \ldots, N_{\varphi_k} - 1$, we denote
them by $\{\Upsilon_{\!(\varphi_k,j)}(m)\}$, periodic in $m$ modulo
$d$ \cite{KBW-GK-I,q-Fourier}; and we assume that they are complete
in ${\mathbb{C}}^d$ and thus have a dual basis
$\{\hat{\Upsilon}_{\!(\varphi_k,j)}(m)\}$ periodic in $m$, such that
\begin{eqnarray}
    \sum_{m=0}^{d-1}
    \hat{\Upsilon}_{\!(\varphi_{k},j)}(m)\,
            \Upsilon_{\!(\varphi_{k'},j')}(m)
        =\delta_{k,{k'}}\delta_{j,j'}
        \label{Upsi-orthocomp1}
\end{eqnarray}
and
\begin{eqnarray}
    \sum_{j=0}^{N_{\varphi_k} - 1}
            \Upsilon_{\!(\varphi_{k},j)}(m)\,
        \hat{\Upsilon}_{\!(\varphi_{k},j)}(m') = ({\bf \Pi}_{\varphi_k})_{mm'}
                \label{Upsi-orthocomp2}
\end{eqnarray}
where ${\bf \Pi}_{\varphi_k}$ is the projector
matrix on the Fourier subspace $\varphi_k$. Associated with this basis
$\{\Upsilon\}$, one may define the corresponding
`$\Upsilon$-fractionalized DFT matrices' ${\bf F_{\!\Upsilon}^\nu}$
with elements
\be
    ({\bf F_{\!\Upsilon}^\nu})_{mm'} :=
        \sum_{k=0}^3\sum_{j=0}^{N_{\varphi_k}-1}
            \Upsilon_{\!(\varphi_{k},j)}(m)\,
                e^{\ii \pi (4j+k) \nu / 2} \,
                    \hat{\Upsilon}_{\!(\varphi_{k},j)}(m')
                        \label{upsilon-DFTmat}
\ee
where we use the compound index $n=4j+k$ to order the vectors, as
if it were the `energy' label in the Mehta functions (\ref{Mehta-func}).
In this way, the vectors of the $\{ \Upsilon \}$ basis are eigenvectors of
a {\em number} matrix ${\bf N_{\Upsilon}}$ with elements
\begin{eqnarray}
    ({\bf N_{\Upsilon}})_{mm'} := \sum_{k=0}^3\sum_{j=0}^{N_{\varphi_k}-1}
            \Upsilon_{\!(\varphi_{k},j)}(m)\,
                (4j{+}k)\,\hat{\Upsilon}_{\!(\varphi_{k},j)}(m')
                    \label{N-Upsi1}
\end{eqnarray}
In other words
\be
    {\bf N_{\Upsilon}} \, {\bf\Upsilon}_{\!(\varphi_{k},j)}
                       = n\,{\bf\Upsilon}_{\!(\varphi_{k},j)}
                    \label{N-Upsi2}
\ee
The matrix ${\bf N_{\Upsilon}}$ has the virtue of being the generator of the
$\Upsilon$-fractional Fourier matrices,
\be
    {\bf F_{\!\Upsilon}^\nu} = \exp(\ii\onehalf\pi\nu\,{\bf N_\Upsilon})
            \label{FracF-expN}
\ee
for $\nu$ modulo 4.

\end{document}